\documentclass[english,twocolumn,conference]{IEEEtran}
\usepackage[T1]{fontenc}
\usepackage{amsmath}
\usepackage{amssymb}
\usepackage{graphicx}

\makeatletter

\providecommand{\tabularnewline}{\\}

\IEEEoverridecommandlockouts
\usepackage{ifpdf}
\usepackage{cite}
\ifCLASSOPTIONcompsoc
\usepackage[caption=false,font=normalsize,labelfont=sf,textfont=sf]{subfig}
\else
\usepackage[caption=false,font=footnotesize]{subfig}
\fi
\usepackage{amssymb}
\usepackage{amsmath}
\usepackage{flushend}

\makeatother

\usepackage{babel}
\begin{document}

\title{Large-scale MU-MIMO: It Is Necessary to Deploy Extra Antennas at
Base Station}

\author{Wei~Ding, Tiejun~Lv,~\IEEEmembership{Senior~Member,~IEEE,}\\
Key Laboratory of Trustworthy Distributed Computing and Service, Ministry
of Education\\School of Information and Communication Engineering
\\ Beijing University of Posts and Telecommunications, Beijing, China
100876\\tepidwater@bupt.edu.cn%
\thanks{This work is financially supported by the National Natural Science
Foundation of China (NSFC) under Grant No. 61271188.%
}}
\maketitle
\begin{abstract}
In this paper, the large-scale MU-MIMO system is considered where
a base station (BS) with extremely large number of antennas ($N$)
serves relatively less number of users ($K$). In order to achieve
largest sum rate, it is proven that the amount of users must be limited
such that the number of antennas at the BS is preponderant over that
of the antennas at all the users. In other words, the antennas at
the BS should be excess. The extra antennas at the BS are no longer
just an optional approach to enhance the system performance but the
prerequisite to the largest sum rate. Based on this factor, for a
fixed $N$, the optimal $K$ that maximizes the sum rate is further
obtained. Additionally, it is also pointed out that the sum rate can
be substantially improved by only adding a few antennas at the BS
when the system is $N=KM$ with $M$ denoting the antennas at each
user. The derivations are under the assumption of $N$ and $M$ going
to infinity, and being implemented on different precoders. Numerical
simulations verify the tightness and accuracy of our asymptotic results
even for small $N$ and $M$.
\end{abstract}

\section{Introduction}

Large-scale multiuser multiple-input multiple-output (LS MU-MIMO)
systems are currently regarded as a novel communication architecture.
By exploiting extremely large number of antennas ($N$) at the base
bastion (BS) to serve relatively less number of users ($K$), several
attractive advantages are emerged, such as the increased system capacity,
the reduced power consumption, and the improved spectral efficiency\cite{howmany,Efficiency,ScalingUp}.
Therefore, the extra antennas at the BS are regarded as an optional
approach to enhance the system performance.

However, with such large transmit dimensions (created by $N$), it
is intuitional to serve the same scale of users for larger sum rate.
But, as shown in this paper, the sum rate decreases instead of increasing
with $K$ if the transmit power at the BS is not sufficiently large
after a certain number of users. Consequently, it is requisite to
keep the transmit dimensions excess for the largest sum rate as well.

This fact is caused by the existence of the multiuser interference
(MUI). Since that the optimal approach to pre-cancel the MUI, the
dirty-paper coding (DPC), is too complex to be implemented, simple
linear precoders are chosen as the only option for LS MU-MIMO systems
\cite{ScalingUp}. While, those linear precoders consumes the transmit
dimensions when they null out the MUI. Hence, the more users are served,
the less transmit dimensions for each user are left. As a result,
if the transmit power is not large enough to compensate for the loss
of transmit dimensions for each user, the sum rate is declined.

Hence, unlike the previous works treating the extra transmit dimensions
as an optional approach to improve the system performance, our contributions
in this paper are that the excess of antennas at the BS is proven
to be a prerequisite to the largest sum rate as well.

More specifically, if $N$ is fixed, it is proven that the largest
sum rate occurs at a certain $K$ where $N$ is greater than the amount
of antennas at all $K$ users. Subsequently, the optimal $K$ that
maximizes the sum rate is further obtained for a given $N$. Though
similar works can be found in \cite{LinearGrowth} and \cite{LinearPrecoding},
their works are only focused on the zero-forcing (ZF) precoders for
LS MU multiple-input single-output (LS MU-MISO) systems where $N$
is much larger than $K$, which can be included as a special case
of our works. Additionally, it is shown that making the system in
$N=KM$ with $M$ denoting the antennas at each user is always not
the optimal strategy, the sum rate can be substantially improved by
only adding a few antennas at the BS. The derivations are under the
assumption of large-scale systems, and being implemented on the ZF
precoders, the regularized ZF (RZF) precoders, and the singular value
decomposition (SVD)-based precoders, respectively. As shown by numerical
simulations, the our results are proven to be asymptotically tight
and accurate for the systems with realistic dimensions.

\noindent \textbf{Notions}: Throughout this paper, vectors and matrices
are denoted by boldface letters. $\left(\cdot\right)^{H}$, $\left(\cdot\right)^{\dagger}$,
$\mathrm{tr}\left(\cdot\right)$, $\mathrm{E}\left[\cdot\right]$,
and $\left\lfloor \cdot\right\rfloor $ denote conjugate transposition,
pseudo-inversion, trace, the expectation, and round down operation,
respectively. Furthermore, `$a.s.$' means almost surely, and $\left(x\right)^{+}=\max\left(0,x\right)$.
The $n$-dimension identity matrix is donated as $\boldsymbol{\mathrm{I}}_{n}$.
$\log$ denotes the logarithm to the base of $2$.

\section{System Model}

\begin{figure}[t]
\begin{centering}
\includegraphics[scale=0.4]{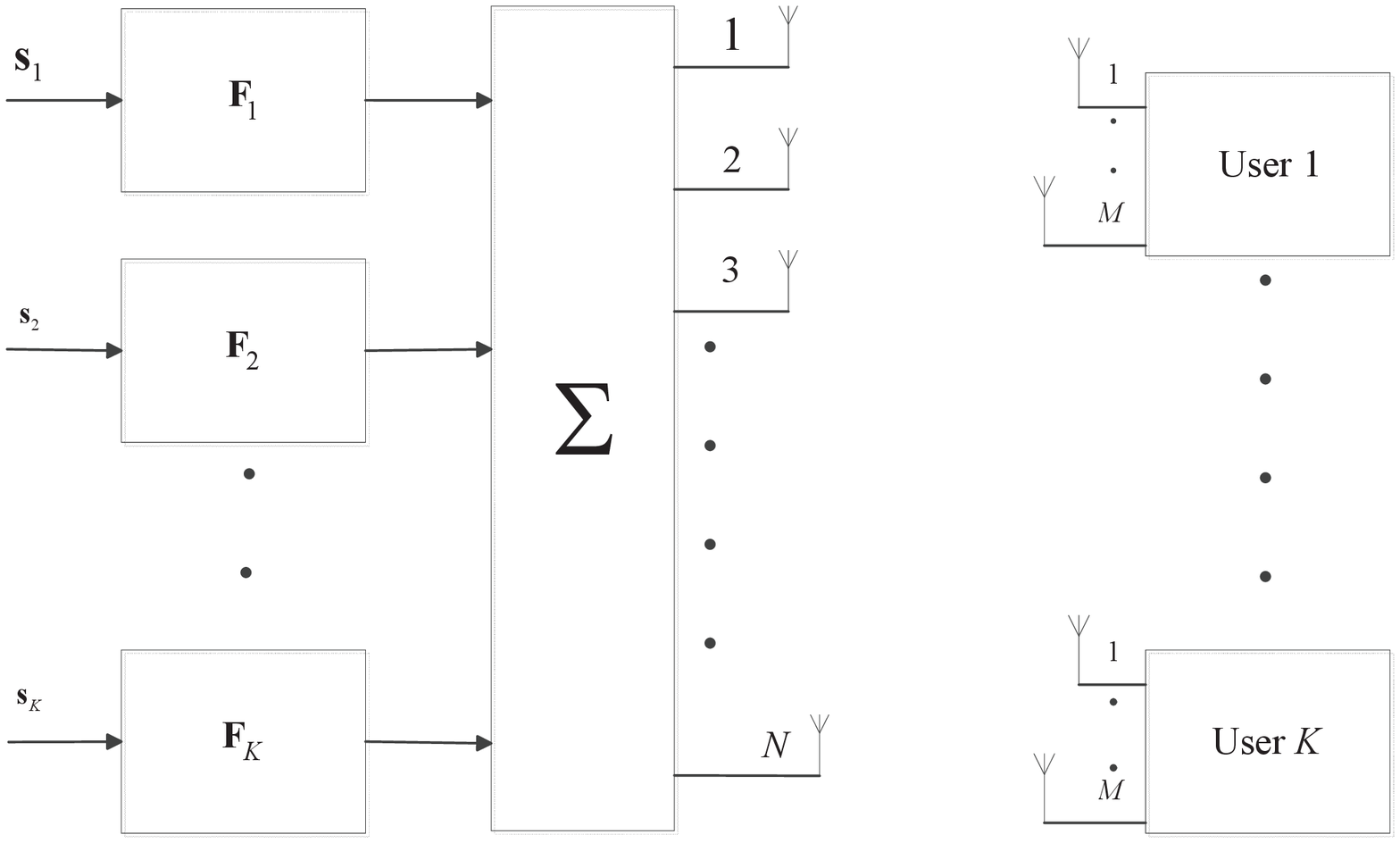}
\par\end{centering}

\caption{\label{fig:1}System model for LS MU-MIMO system}

\end{figure}

\global\long\def\matrix#1#2{\boldsymbol{\mathrm{#1}}_{#2}}
\global\long\def\exception#1{\mathrm{E}\left[#1\right]}
\global\long\def\trace#1{\mathrm{tr}\left(#1\right)}
\global\long\def\mat#1{\boldsymbol{\mathrm{#1}}}
Consider a single-cell downlink MU-MIMO system in Fig. \ref{fig:1},
 which comprises of a central BS with $N$ antennas and $K$ uncooperative
users with $M$ antennas. $N\geq KM$ is preferred, thus, user scheduling
is not taken into account.The received signal for the $k$-th user
is given by
\begin{equation}
\matrix yk=\matrix Hk\matrix Fk\matrix sk+\matrix Hk\sum_{l\neq k}^{K}\matrix Fl\matrix sk+\matrix nk,\label{eq:1}
\end{equation}
 where $\matrix yk\in\mathbb{C}^{M\times1}$, $\matrix sk\in\mathbb{C}^{M\times1}$,
$\matrix Hk\in\mathbb{C}^{M\times N}$, $\matrix Fk\in\mathbb{C}^{N\times M}$,
and $\matrix nk\in\mathbb{C}^{M\times1}$ are the received signal,
information-bearing signal, channel matrix, precoder matrix, and Gaussian
thermal noise vector for the $k$-th user, respectively. Since a LS-MIMO
system is assumed, the entries of $\matrix Hk$ can be modeled as
the identically independently distributed (i.i.d) circularly-symmetric
complex Gaussian distribution with zero mean and variance $\sigma_{n}^{2}$,
namely, $\mathcal{CN}\left(0,\sigma_{n}^{2}\right)$. For frequency
selective fading channels, this model can be extended by using orthogonal
frequency division multiplexing modulation (OFDM) \cite{LinearGrowth}.
The channel side information at transmitter (CSIT) is assumed.

Clearly, the transmit dimensions for the $k$-th user is $N$, but
the current user suffers the MUI, namely, the second term of the right
hand side (RHS) of \eqref{eq:1}. Hence, the precoding techniques
are required to be utilized at the BS to pre-cancel the MUI.

The linear precoder are designed for single-user (SU) MIMO (SU-MIMO)
systems, which can be extended to MU-MIMO systems by block diagonalization
(BD) technique \cite{BD}. The precoder for BD technique is a cascade
of two precoding matrices, namely, $\matrix Fk=\matrix Bk\matrix Dk$,
where $\matrix Bk$ removes the MUI and $\matrix Dk$ can be further
designed under the different criteria. To do so, $\matrix Bk$ should
be chosen from the null space of $\matrix Hl$ ($\forall l\neq k$),
i.e., $\matrix Hl\matrix Bk=\mat 0$ for any $l\neq k$. In particular,
if $\tilde{\mat H}_{k}$ is defined as $\tilde{\mat H}_{k}=\left[\boldsymbol{\mathrm{H}}_{1}^{H},\dots,\mat H_{k-1}^{H},\mat H_{k+1}^{H},\dots,\mat H_{K}^{H}\right]^{H}$,
then $\matrix Bk$ can be obtained by the SVD on $\tilde{\mat H}_{k}$,
namely,
\[
\tilde{\mat H}_{k}=\tilde{\mat U}_{k}\left[\tilde{\mat{\Sigma}}_{k}\;\mat 0\right]\left[\tilde{\mat V}_{k}^{\left(1\right)}\;\tilde{\mat V}_{k}^{\left(0\right)}\right]^{H},
\]
 where $\tilde{\mat U}_{k}$ and $\tilde{\mat{\Sigma}}_{k}$ are the
left singular vector matrix and the matrix of ordered singular values
of $\tilde{\mat H}_{k}$, respectively. Matrices $\tilde{\mat V}_{k}^{\left(1\right)}$
and $\tilde{\mat V}_{k}^{\left(0\right)}$ denote the right singular
matrices, each of them consists of the singular vectors corresponding
to non-zero singular values and zero singular values of $\tilde{\mat H}_{k}$,
respectively. Note that $\matrix Hl\tilde{\mat V}_{k}^{\left(0\right)}=\mat 0$
($\forall l\neq k$), $\matrix Bk$ is obtained by choosing $L_{k}$
columns from the $\tilde{\mat V}_{k}^{\left(0\right)}$. To ensure
there are at least $L_{k}$ columns in each $\tilde{\mat V}_{k}^{\left(0\right)}$,
$L_{k}$ should satisfy the dimensionality constraint as
\begin{equation}
L_{k}\leq N-\left(K-1\right)M.\label{eq:2}
\end{equation}

\begin{figure}[t]
\begin{centering}
\includegraphics[scale=0.45]{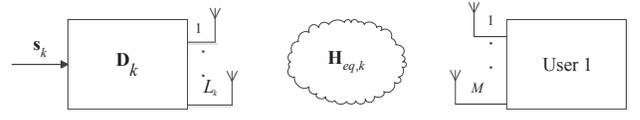}
\par\end{centering}

\caption{\label{fig:2}Equivalent system for each user}

\end{figure}

Therefore, with $\matrix Bk$, \eqref{eq:1} is rewritten as
\begin{equation}
\matrix yk=\matrix H{eq,k}\matrix Dk\matrix sk+\matrix nk,\label{eq:3}
\end{equation}
 where $\matrix H{eq,k}=\matrix Hk\matrix Bk\in\mathbb{C}^{M\times L_{k}}$
is the equivalent channel for $k$-th user. The transmission model
in \eqref{eq:3} can be interpreted as an equivalent SU-MIMO system
in Fig. \ref{fig:2},
 which a BS with $L_{k}$ antennas communicates\emph{ }with a receiver
with $M$ antennas. $\matrix Dk\in\mathbb{C}^{L_{k}\times M}$ is
therefore the precoder for such system. To ensure the existence of
$\matrix Dk$, $L_{k}\geq M$ needs to be satisfied too.

Obviously, \eqref{eq:3} does not suffer the MUI, but it is worth
noticing that the transmit dimensions for each user have been declined
from $N$ in \eqref{eq:1} to $L_{k}$ in \eqref{eq:3}. The loss
of transmit dimensions will definitely decrease the rate for the current
user. However, due to the cancellation of the MUI, the entire LS MU-MIMO
system achieves the full degrees of freedom (DoFs) promised by the
DPC precoding. If the transmit power at the BS can enhance unlimitedly
to compensate for the loss of transmit dimensions, the sum rate will
increase with $K$. While, for the case of limited transmit power,
there is tradeoff between the number of users and the transmit dimensions.

Define the sum rate of the MU-MIMO system using BD technique as
\begin{equation}
\mathcal{R}_{sum}=\sum_{k=1}^{K}\mathcal{R}_{k}\label{eq:4}
\end{equation}
 with
\[
\mathcal{R}_{k}=\mathrm{E}\left\{ \log\det\left(\matrix I{M_{k}}+\frac{1}{\sigma_{n}^{2}}\matrix H{eq,k}\matrix Q{eq,k}\mat H_{eq,k}^{H}\right)\right\} ,
\]
 where $\matrix Q{eq,k}=\exception{\mat D_{k}\mat D_{k}^{H}}$ denotes
the input covariance matrix and the $\sigma_{n}^{2}$ is the variance
of the noise. Since the transmit power is always limited, the following
works in this paper is to quantify the impact of the tradeoff on $\mathcal{R}_{sum}$.

\section{Asymptotic Sum Rate of Different Precoders}

In this section, the asymptotic sum rate performance of three different
precoders is derived, as a groundwork of analysis in the next section.
Before further introducing current section, a basic theorem is given
at first.

\textbf{Theorem}: Consider two random matrices $\mat A\in\mathbb{C}^{M\times N}$
and $\mat B\in\mathbb{C}^{N\times L}$, where the entries in $\mat A$
follow i.i.d $\mathcal{CN}\left(0,1\right)$ and $\mat B^{H}\mat B=\matrix IL$.
If $\mat A$ is independent of $\mat B$, the entries in $\mat H=\mat A\mat B$
share the same distributions as those in $\mat A$.
\begin{IEEEproof}
\noindent See in \cite{BD}, also in \cite{BD3}.
\end{IEEEproof}

\subsection{Singular Value Decomposition-based Precoder}

The principle of SVD-based precoder is to decompose the system channel
into several parallel sub-channels, then implementing the power allocation
using water-filling algorithm to make all the exploited sub-channels
have the same gains.

Let $\mat D_{k}=\mat V_{k}\mat M_{k}^{1/2}$, where $\mat V_{k}$
is the right singular matrix of $\mat H_{eq,k}$ and $\matrix Mk$
is the diagonal power allocation matrix. Assuming the transit power
allocated to the $k$-th user  is $P_{k}$, The principle of SVD-based
precoder can be mathematically expressed as
\[
\mathcal{R}_{k}^{SVD}=\mathrm{E}\left\{ \sum_{i=1}^{M}\left[\log\left(M\nu\lambda_{i}\right)\right]^{+}\right\}
\]
 with the power level $\nu$ obtained from
\[
\frac{1}{M}\sum_{i=1}^{M}\left(M\nu-\frac{1}{\lambda_{i}}\right)^{+}=\rho_{k},
\]
 where $\lambda_{i}$ denotes the $i$-th eigenvalue of $M^{-1}\mat H_{eq,k}\mat H_{eq,k}^{H}$
and $\rho_{k}=P_{k}/\sigma_{n}^{2}$.

Clearly, $\mathcal{R}_{k}^{SVD}$ depends on $\lambda_{i}$ ($i=1,\dots,M$).
Since the distributions of eigenvalues for a matrix with regular dimensions
are too complex to be further analyzed, the asymptotic behavior of
$\lambda_{i}$ is mainly focused.

Based on the Theorem above, the entries in $\mat H_{eq,k}$ follow
the i.i.d zero-mean Gaussian distribution with unity variance. Thus,
when $\rho_{k}\geq\frac{2}{\left(\sqrt{\beta_{k}}-1\right)\left(\beta_{k}-1\right)}$,
$\mathcal{R}_{k}^{SVD}$ almost surely converges to a deterministic
value as $L_{k},M\rightarrow\infty$ with $L_{k}/M\rightarrow\beta_{k}$
\footnote{ It should be noticed that $N$ is scaling up correspondingly to hold
the dimension constraint in \eqref{eq:2}.%
}, namely,
\begin{multline}
\mathcal{R}_{k}^{SVD}\xrightarrow{a.s.}M\Big[\log\left(1+\rho_{k}\left(\beta_{k}-1\right)\right)+\beta_{k}\log\beta_{k}\\
-\beta_{k}\log\left(\beta_{k}-1\right)-\log e\Big],\label{eq:5}
\end{multline}
 where the derivation is directly based on \cite{SVD} and the constraint
on $\rho_{k}$ ensures all the sub-channels are exploited for data
transmission.

\eqref{eq:5} only holds in the case of $\beta_{k}>1$. For $\beta_{k}=1$,
the minimum eigenvalue of $M^{-1}\mat H_{eq,k}\mat H_{eq,k}^{H}$,
denoted by $\lambda_{\min}$, converges to zero under large-scale
system assumption \cite{eig}. In this case, only part of the the
sub-channels will be used. Under the circumstance, $\mathcal{R}_{k}^{SVD}$
shows almost sure convergence to
\begin{equation}
\mathcal{R}_{k}^{SVD}\xrightarrow{a.s.}\int_{\bar{\nu}^{-1}}^{4}\log_{2}\left(\bar{\nu}x\right)f_{1}\left(x\right)dx\label{eq:6}
\end{equation}
 with $\bar{\nu}=M\nu$ calculated from
\begin{equation}
\int_{\bar{\nu}^{-1}}^{4}\left(\bar{\nu}-\frac{1}{x}\right)f_{1}\left(x\right)dx=\rho_{k},\label{eq:7}
\end{equation}
 where $f_{1}\left(x\right)$ is the probability density of Mar\u{c}enko-Pastur
distribution with parameter one \cite{RMT}. Solving the integral
in \eqref{eq:7} yields
\begin{equation}
\rho_{k}=\frac{1}{2\pi}\left[(1+2\bar{\nu})\arccos\left(\frac{1-2\bar{\nu}}{2\bar{\nu}}\right)-3\sqrt{4\bar{\nu}-1}\right].\label{eq:8}
\end{equation}
 Since the right hand side (RHS) in \eqref{eq:8} is a monotonically
increasing function of $\bar{\nu}$ in the region $\left[0.25,\infty\right)$,
the solution is unique for every positive value of $\rho_{k}$. Subsequently,
\eqref{eq:6} can be calculated using numerical integral technique.

Similarly to the definition in \eqref{eq:4}, the asymptotic sum rate
of SVD-based precoder, $\bar{\mathcal{R}}_{sum}^{SVD}$, can be obtained
via \eqref{eq:5} and \eqref{eq:6} for different values of $\beta_{k}$.

\subsection{Zero-forcing Precoder}

Unlike the SVD-based precoder decomposing the entire channel into
several sub-channels, the ZF precoder `smoothes' the channel by inversion
operation, hence, the received signal is just a scaled version of
information-bearing signal. Specifically, by defining $\matrix Dk=\kappa\mat H_{eq,k}^{\dagger}$,
the rate of $k$-th UT, $\mathcal{R}_{k}^{ZF}$, is revised as
\[
\mathcal{R}_{k}^{ZF}=M\mathrm{E}\left\{ \log\left(1+\kappa^{2}/\sigma_{n}^{2}\right)\right\} ,
\]
 in which $\kappa=\sqrt{P_{k}/\mathrm{tr}\left(\matrix Dk\mat D_{k}^{H}\right)}$
is the power constraint factor. Clearly, $\kappa$ dominants the performance
of $\mathcal{R}_{k}^{ZF}$. To have a friendly expression, the $\mathcal{R}_{k}^{ZF}$
is rewritten as
\begin{eqnarray}
\mathcal{R}_{k}^{ZF} & = & M\mathrm{E}\left\{ \log\left(1+\rho_{k}/\mathrm{tr}\left[\left(\matrix H{eq,k}\mat H_{eq,k}^{H}\right)^{-1}\right]\right)\right\} ,\nonumber \\
 & \xrightarrow{a.s.} & M\mathrm{E}\left\{ \log\left(1+\rho_{k}\left(\beta_{k}-1\right)\right)\right\} ,\label{eq:9}
\end{eqnarray}
 where the `$\xrightarrow{a.s.}$' abides by the asymptotic behavior
of $M^{-1}\mat H_{eq,k}\mat H_{eq,k}^{H}$ \cite{LinearGrowth}.

Obviously, \eqref{eq:9} does not hold when $\beta_{k}$ approaches
one. That is because that the $\lambda_{\min}$ will converge to zero
with probability one in the case of $\beta_{k}=1$, the first order
moment of $\left(\matrix H{eq,k}\mat H_{eq,k}^{H}\right)^{-1}$ is
no longer converged.

In order to evaluate the asymptotic performance of $\mathcal{R}_{k}^{ZF}$
when $\beta_{k}$ equals one, the upper bound of $\mathcal{R}_{k}^{ZF}$
is derived, namely,
\begin{eqnarray}
\mathcal{R}_{k}^{ZF} & = & M\mathrm{E}\left\{ \log\left(1+\rho_{k}\left(M^{-1}\sum_{i=1}^{M}\lambda^{-1}\right)^{-1}\right)\right\} ,\nonumber \\
 & \leq & \mathrm{E}\left\{ \sum_{i=1}^{M}\log\left(1+\rho_{k}\lambda_{i}\right)\right\} ,\label{eq:10}
\end{eqnarray}
 where `$\leq$' follows from Jensen's inequality. The RHS of \eqref{eq:10}
converges to the following integral with probability one as $L_{k},M\rightarrow\infty$
with $L_{k}/M\rightarrow\beta$, i.e.,
\[
\mathrm{E}\left\{ \sum_{i=1}^{M}\log\left(1+\rho_{k}\lambda_{i}\right)\right\} \xrightarrow{a.s.}M\int\log\left(1+\rho_{k}x\right)f_{\beta_{k}}\left(x\right)dx,
\]
 in which $f_{\beta_{k}}\left(x\right)$ is the probability density
of Mar\u{c}enko-Pastur distribution with parameter $\beta_{k}$.
The integral yields a closed-form solution, which is
\begin{multline}
\mathcal{R}_{k}^{ZF}\preccurlyeq M\Big[\beta_{k}\log\left(1+\rho_{k}-\mathcal{F}\left(\rho_{k},\beta_{k}\right)\right)\\
+\log\left(1+\rho_{k}\beta_{k}-\mathcal{F}\left(\rho_{k},\beta_{k}\right)\right)-\frac{\log e}{\rho_{k}}\mathcal{F}\left(\rho_{k},\beta_{k}\right)\Big],\label{eq:11}
\end{multline}
where
\[
\mathcal{F}\left(x,y\right)=\left(\sqrt{x\left(1+\sqrt{y}\right)^{2}+1}-\sqrt{x\left(1-\sqrt{y}\right)^{2}+1}\right)^{2}
\]
 and `$\preccurlyeq$' denotes asymptotically equal or less than.

Consequently, the asymptotic sum rate of ZF precoder, $\bar{\mathcal{R}}_{sum}^{ZF}$,
can be derived from \eqref{eq:9} and \eqref{eq:11} for by \eqref{eq:4}
for different values of $\beta_{k}$.

\subsection{Regularized Zero-forcing Precoder}

The RZF precoder is proposed to improve the performance of ZF precoder
in the case of $\beta_{k}=1$ by adding a multiple of the identify
matrix before inversion, namely, $\matrix Dk=\kappa\mat W\mat H_{eq,k}^{H}$
with $\mat W=\left(\mat H_{eq,k}^{H}\mat H_{eq,k}+M\rho_{k}^{-1}\matrix I{L_{k}}\right)^{-1}$.

Unlike the previous precoders, the received signal suffers the inter-stream
interference. Recalling the transmission model in \eqref{eq:3}, the
$i$-th stream for the $k$-th user is given as
\[
y_{i}=\kappa\matrix hi\mat W\mat h_{i}^{H}s_{i}+\kappa\matrix hi\sum_{j\neq i}^{M_{k}}\mat W\mat h_{j}^{H}s_{j}+n_{i},
\]
 where $\matrix hi$ $y_{i}$, $s_{i}$, and $n_{i}$ denote the $i$-th
row of $\matrix H{eq,k}$, the $i$-th elements in $\matrix yk$,
$\matrix sk$, and $\matrix nk$, respectively. Hence, $\mathcal{R}_{k}^{RZF}$
is redefined as a function of received signal-to-noise-plus-interference-ratio
(SINR) for each stream, i.e.,
\[
\mathcal{R}_{k}^{RZF}=\mathrm{E}\left\{ \sum_{i=1}^{M}\log\left(1+\gamma_{i}\right)\right\}
\]
with the SINR for the $i$-th stream
\[
\gamma_{i}=\frac{\left|\mat h_{i}\mat W\mat h_{i}^{H}\right|^{2}}{\left(\mat h_{i}\mat W\mat H_{\left[i\right]}^{H}\mat H_{\left[i\right]}\mat W\mat h_{i}+\sigma_{n}^{2}/\kappa^{2}\right)},
\]
 where $\mat H_{\left[i\right]}$ is the $\mat H_{eq,k}$ with $i$-th
row removed.

Before introducing the asymptotic behavior of $\mathcal{R}_{k}^{RZF}$,
it is pointed out that the largest eigenvalue of $M^{-1}\mat H_{eq,k}\mat H_{eq,k}^{H}$,
$\lambda_{\max}$, shows the almost sure convergence to a deterministic
value as $L_{k},M\rightarrow\infty$ with $L_{k}/M\rightarrow\beta_{k}$,
namely \cite{eig},
\[
\lambda_{\max}\xrightarrow{a.s.}\left(1+\sqrt{\beta_{k}}\right)^{2}.
\]
 Therefore, $M^{-1}\mat H_{eq,k}\mat H_{eq,k}^{H}$ has uniformly
bounded spectral norm on $M$ with probability one. Based on this
fact, $\mathcal{R}_{k}^{RZF}$ converges to the following equation
with the large-scale system assumption, i.e.,
\begin{multline}
\hspace{1em}\hspace{-5mm}\mathcal{R}_{k}^{RZF}\xrightarrow{a.s.}M\Bigg[\log_{2}\Bigg(1+\rho_{k}\left(\beta_{k}-1\right)\\
+\sqrt{\rho_{k}^{2}\left(\beta_{k}-1\right)^{2}+2\rho_{k}\left(\beta_{k}+1\right)+1}\Bigg)-1\Bigg],\label{eq:12}
\end{multline}
 which is based on \cite[\textit{Corollary 2}]{LinearPrecoding}.

Consequently, the asymptotic sum rate of RZF precoder, $\bar{\mathcal{R}}_{sum}^{RZF}$,
is achievable by substituting \eqref{eq:12} into \eqref{eq:4}.

\section{Results and Numerical Simulations}

In this section, the main results in this paper are presented, and
being verified by numerical simulations. There is tradeoff between
transmit dimensions and the number of users is given at first, followed
by the optimal $K$ that maximizes the $\mathcal{R}_{sum}$.

Throughout this section, a $\left(N,M,K\right)$ system is referred
as a $K$ users system with $N$ transmit antennas at the BS and $M$
receive antennas at each user. $L_{k}$ and $P_{k}$ are assumed to
be $L=N-\left(K-1\right)M$ and $P$ for all the users. The assumption
is logical, because the large transmit dimensions maximizes the rate
per user, and it is unnecessary to implement power allocation among
statistically identical users. Additionally, `ZF', `RZF', and `SVD'
are abbreviations for the MU-MIMO systems using BD technique with
ZF precoder, RZF precoder, and SVD-based precoder, respectively.

Based on the configurations above, $\beta_{k}$ is further rewritten
as
\begin{equation}
\beta_{k}=\beta-K+1,\label{eq:13}
\end{equation}
 where $\beta=N/M$ is the normalized the transmit dimensions. And
$\rho_{k}$ is set to $\rho$ for all the user with $\rho=P/\sigma_{n}^{2}.$
Clearly, the normalized transmit dimensions per user ($\beta_{k}$)
is declined with $K$.

To quantify the impact of the tradeoff on the sum rate, the asymptotic
sum rates of above three different precoders is summarized into an
unified form, donated by $\bar{\mathcal{R}}_{sum}$, namely,
\begin{equation}
\bar{\mathcal{R}}_{sum}\left(K,\rho\right)=KM\left[\mathcal{I}_{1}\left(K,\rho\right)+\mathcal{I}_{2}\left(K,\rho\right)\right],\;\text{for }K<\beta\label{eq:14}
\end{equation}
 where $\mathcal{I}_{1}\left(K,\rho\right)=\log\left(1+\rho\left(\beta-K\right)\right)$,
and $\mathcal{I}_{2}\left(K,\rho\right)$ varies from different precoders
whose explicit forms can be found in Table \ref{tab:t1}.
\begin{table}[t]
\caption{\label{tab:t1}Explicit forms of array gain for different precoders}

\centering{}%
\begin{tabular}{|c|c|}
\hline
 & $\mathcal{I}_{2}\left(K,\rho\right)$\tabularnewline
\hline
\hline
SVD & $\left(\beta-K+1\right)\log\left(1+\frac{1}{\beta-K}\right)+\log e$\tabularnewline
\hline
RZF & $\log\left(1+\sqrt{1+\frac{4\rho}{\left(\rho\left(\beta-K\right)+1\right)^{2}}}\right)-1$\tabularnewline
\hline
ZF  & 0\tabularnewline
\hline
\end{tabular}
\end{table}
 For the case of $K=\beta$, $\bar{\mathcal{R}}_{sum}\left(K,\rho\right)$
can be derived from \eqref{eq:4}, \eqref{eq:6}, \eqref{eq:11} and
\eqref{eq:12} for different precoders.

From \eqref{eq:14}, it is observed that $\bar{\mathcal{R}}_{sum}\left(K,\rho\right)$
can be decomposed into three terms, which are promised by the number
of users ($KM$), the excess of transmit dimensions per user ($\mathcal{I}_{1}\left(K,\rho\right)$),
and array gain ($\mathcal{I}_{2}\left(K,\rho\right)$), respectively%
\footnote{The analysis for the case of $K=\beta$ is omitted, because that the
number of users of that case is fixed.%
}. Those three terms impact on the sum rate differently. The number
of users dominants the DoFs, which is ratio of $\mathcal{\bar{R}}_{sum}$
increasing with $\rho$. While the last two terms determine the `starting
point' from which $\bar{\mathcal{R}}_{sum}$ increases with $\rho$.

When $K$ increases, on one hand, the DoFs is definitely increased.
On the other hand, the transmit dimensions are declined, which leads
to an ill-conditioned equivalent channel matrix for each user. Therefore,
the second term is decreased when $K$ increases. But since the array
gain is benefited from the coherent combining of channel gains, the
larger the condition number of each equivalent channel is, the more
array gain each user is obtained in the presence of CSIT%
\footnote{Notice that ZF omits the channel gains by the inversion operation,
hence, it can not obtain the array gain. Besides, RZF develops into
ZF at high SNR regime, it explains why the array gain vanishes as
$\rho\rightarrow\infty$.%
} \cite{WC}. Hence, same to the first term, the third term increases
with $K$. However, because of the properties of $\log\left(1+x\right),$
the decrement of $\mathcal{I}_{1}\left(K,\rho\right)$ exceeds the
increment of $\mathcal{I}_{2}\left(K,\rho\right)$. As a result, the
`starting point' of $\bar{\mathcal{R}}_{sum}$ increasing with $\rho$
is declined when $K$ gets large. In conclusion, the tradeoff between
$K$ and the transmit dimensions per user can be interpreted as the
tradeoff between the ratio and `starting point' of $\bar{\mathcal{R}}_{sum}$
increasing with $\rho$ .

According to above analysis, it is intuitional that the highest sum
rate is achieved by serving users as many as possible when the transmit
power is unlimited. But, when the transmit is not sufficient large,
the sum rate of a system with more users may be lower than that of
a system with relatively less users. It is because that the sum rate
with higher increasing ratio has not catched up with those rising
from a higher `starting point' yet at the given $\rho$.

\begin{figure}[t]
\begin{centering}
\includegraphics[scale=0.45]{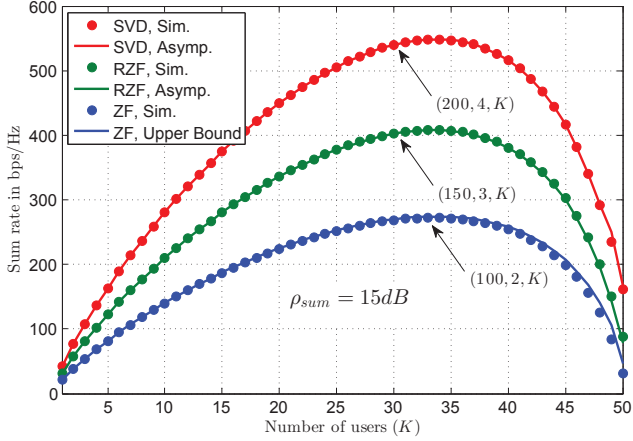}
\par\end{centering}

\caption{\label{fig:3}Sum rate performance for different systems with different
precoders}
\end{figure}

Before further finding the optimal number of users, the accuracy and
tightness of $\bar{\mathcal{R}}_{sum}\left(K,\rho\right)$ are verified
via numerical simulation in Fig. \ref{fig:3},
 where $\rho_{sum}$ denotes the total transmit SNR for all users.
`Sim.' and `Aysmp.' are abbreviations for simulation and asymptotic
results, respectively. As shown in Fig.\ref{fig:3}, our asymptotic
sum rates and upper bound for $\mathcal{R}_{sum}^{ZF}$ are very tight
even for small $M$ and $N$. And the sum rates of different precoders
increase with $K$ firstly but decreases after a certain value of
$K$ as analyzed above.

The optimal $K$ where the highest sum rate occurs, denoted by $K^{*}$,
is
\begin{equation}
K^{*}=\arg\max_{K\in\mathcal{K}}\bar{\mathcal{R}}_{sum}\label{eq:16}
\end{equation}
 for a fixed $\rho$, where  $\mathcal{K}$ is the set including all
candidates of $K$, i.e., $\mathcal{K}\in\left\{ 1,\dots,\left\lfloor \beta\right\rfloor \right\} $.
The exact $K^{*}$ involves solving the equations with the form of
$x=\log_{2}\left(1+x\right)$, which can not be concluded into an
explicit form. However, since $\bar{\mathcal{R}}_{sum}\left(\rho,K\right)$
is an unary offline function of $K$ for a given $\rho$, implementing
$1-D$ searching over $\mathcal{K}$ is also a low-complexity solution.

\begin{figure}[t]
\begin{centering}
\includegraphics[scale=0.45]{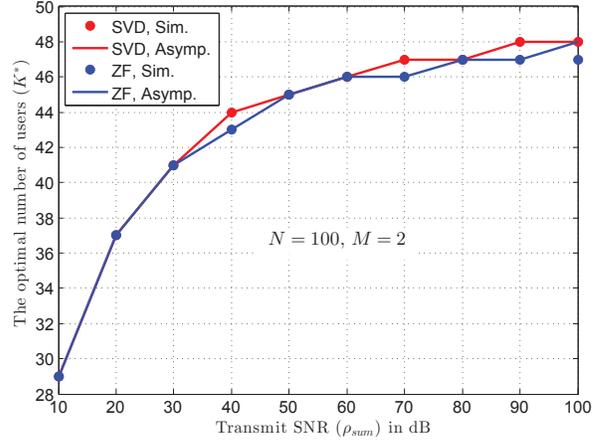}
\par\end{centering}

\caption{\label{fig:4}The optimal user number v.s. transmit SNR}
\end{figure}

The behavior of $K^{*}$ when $\rho$ increases is evaluated in Fig.
\ref{fig:4}.
 Since RZF develops into ZF at high SNR regime, its behavior is omitted
in the Fig. \ref{fig:4}. It is observed that $K^{*}$ increases with
$\rho$, which is corresponding to the previous analysis. The mismatches
of ZF after $\rho_{sum}=90dB$ is likely caused by the usage of upper
bound. It is obvious that serving users as large as possible guarantees
the highest sum rate when the transmit power is unlimited. However,
it also can be seen that $K^{*}$ being the maximum $K$ only happens
at very high $\rho_{sum}$ (clearly should be greater than $100dB$),
which indicates that making the system fully load ($K=\beta$) is
not the best strategy when the transmit power is limited.

\begin{figure}[t]
\begin{centering}
\includegraphics[scale=0.45]{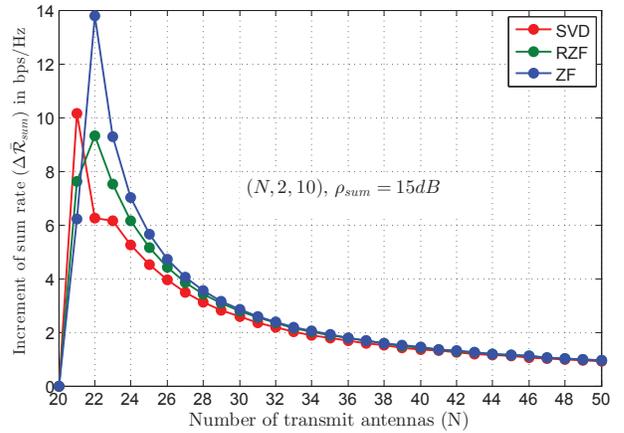}
\par\end{centering}

\caption{\label{fig:5}Increment of asymptotic sum rate brought by adding one
antenna}
\end{figure}

Specifically, $\Delta\bar{\mathcal{R}}_{sum}$, the increment of $\bar{\mathcal{R}}_{sum}$
brought by only adding one antenna at the BS compared to previous
systems, is plotted in Fig. \ref{fig:5}.
 For example, if the original system is a $\left(N,M,K\right)$ system,
$\Delta\bar{\mathcal{R}}_{sum}$ donates the increment of $\bar{\mathcal{R}}_{sum}$
from $\left(N+1,M,K\right)$ to $\left(N,M,K\right)$, or from $\left(N+2,M,K\right)$
to $\left(N+1,M,K\right)$ and so on. In Fig. \ref{fig:5}, the original
system configuration is $\left(20,2,10\right)$, it is clear that
$\bar{\mathcal{R}}_{sum}$ can be substantially improved by only adding
a few antennas at the BS. That is because all the equivalent systems
will have the same number of extra transmit antennas when more antennas
are added at the BS. Hence, the increment of sum rate is multiplied
by the number of users. While, $\Delta\bar{\mathcal{R}}_{sum}$ increases
slowly when $N$ is getting large. Under such circumstance, it is
the MU gain or DoFs that dominants the sum rate. Though the sum rate
can always benefits from more antennas at the BS, it is worth noticing
that $\Delta\bar{\mathcal{R}}_{sum}\rightarrow0$ when $N\rightarrow\infty$,
which can be proven by derivation of $\bar{\mathcal{R}}_{sum}$ with
respect to $\beta$. Therefore, it is unnecessary to endow too many
antennas at the BS by taking the expenses brought by them into consideration.

\section{Conclusion }

In this paper, the tradeoff between MU gain and transmit diversity
for MU-MIMO systems using BD technique is established. Specifically,
a expression is derived to quantify how the two factors impact on
the sum rate. Based on the tradeoff, the optimal number of users that
maximizes the sum rate is obtained as well. The results in this paper
have important significance for system configuration in practice.
Note that the derivations is under the Rayleigh channels, they will
be extended to other channel models in the future works.

\bibliographystyle{IEEEtran}
\bibliography{ciations}

\begin{thebibliography}{10}
\providecommand{\url}[1]{#1}
\csname url@samestyle\endcsname
\providecommand{\newblock}{\relax}
\providecommand{\bibinfo}[2]{#2}
\providecommand{\BIBentrySTDinterwordspacing}{\spaceskip=0pt\relax}
\providecommand{\BIBentryALTinterwordstretchfactor}{4}
\providecommand{\BIBentryALTinterwordspacing}{\spaceskip=\fontdimen2\font plus
\BIBentryALTinterwordstretchfactor\fontdimen3\font minus
  \fontdimen4\font\relax}
\providecommand{\BIBforeignlanguage}[2]{{%
\expandafter\ifx\csname l@#1\endcsname\relax
\typeout{** WARNING: IEEEtran.bst: No hyphenation pattern has been}%
\typeout{** loaded for the language `#1'. Using the pattern for}%
\typeout{** the default language instead.}%
\else
\language=\csname l@#1\endcsname
\fi
#2}}
\providecommand{\BIBdecl}{\relax}
\BIBdecl

\bibitem{howmany}
J.~{Hoydis et. al.}, ``Massive {MIMO}: How many antennas do we need?'' in
  \emph{Proc. 49th Annu. Allerton Conf. on Communication, Control and
  Computing}, Monticello, IL, 2011, pp. 545--550.

\bibitem{Efficiency}
T.~L.~M. Hien Quoc~Ngo, Erik G.~Larsson, ``Energy and spectral efficiency of
  very large multiuser {MIMO} systems,'' \emph{IEEE Trans. Commun.}, to be
  published.

\bibitem{ScalingUp}
F.~Rusek, D.~Persson, B.~K. Lau, E.~G. Larsson, T.~L. Marzetta, O.~Edfors, and
  F.~Tufvesson, ``Scaling up {MIMO}: Opportunities and challenges with very
  large arrays,'' \emph{IEEE Signal Processing Mag.}, vol.~30, no.~1, pp.
  40--60, Jan. 2013.

\bibitem{LinearGrowth}
B.~Hochwald and S.~Vishwannath, ``Space-time multiple access: Linear growth in
  the sum-rate,'' in \emph{Proc. 40th Annu. Allerton Conf. on Communication,
  Control and Computing}, Monticello, IL, 2002.

\bibitem{LinearPrecoding}
S.~Wagner, R.~Couillet, M.~Debbah, and D.~T.~M. Slock, ``Large system analysis
  of linear precoding in correlated {MISO} broadcast channels under limited
  feedback,'' \emph{IEEE Trans. Inf. Theory}, vol.~58, no.~7, pp. 4509--4537,
  Jul. 2012.

\bibitem{BD}
L.~Choi and R.~D. Murch, ``A transmit preprocessing technique for multiuser
  {MIMO} systems using a decomposition approach,'' \emph{IEEE Trans. Wireless
  Commun.}, vol.~3, no.~1, pp. 20--24, Jan. 2004.

\bibitem{BD3}
S.~Shim, J.~Kwak, R.~Heath, and J.~Andrews, ``Block diagonalization for
  multi-user {MIMO} with other-cell interference,'' \emph{IEEE Trans. Wireless
  Commun.}, vol.~7, no.~7, pp. 2671--2681, Jul. 2008.

\bibitem{SVD}
A.~Tulino and A.~Lozano, ``{MIMO} capacity with channel state information at
  the transmitter,'' in \emph{Proc. IEEE ISSTA}, Boston, MA, 2004, pp. 22--26.

\bibitem{eig}
Z.~D. Bai, ``Methodologies in spectral analysis of large dimensional random
  matrices,'' \emph{Statistica Sinica}, vol.~9, pp. 611--661, 1999.

\bibitem{RMT}
A.~M. Tulino and S.~Verd\'{u}, \emph{Random matrix theory and wireless
  communications}.\hskip 1em plus 0.5em minus 0.4em\relax Boston, MA: Now
  Publishers Inc., 2004.

\bibitem{WC}
A.~Goldsmith, \emph{Wireless Communications}.\hskip 1em plus 0.5em minus
  0.4em\relax Cambridge, UK: Cambridge University Press, 2005.

\end{thebibliography}

\end{document}